\documentclass[10pt,conference]{IEEEtran}

\usepackage{graphicx}
\usepackage{amsmath}
\usepackage{booktabs}
\usepackage{hyperref}
\usepackage{pifont}
\usepackage{tcolorbox}
\usepackage{colortbl}
\usepackage{multirow}
\usepackage{xcolor}
\usepackage{xspace}
\usepackage{graphicx}   
\usepackage{caption}    
\usepackage{subcaption}  
\usepackage{booktabs}
\usepackage{graphicx}
\usepackage{enumitem}
\usepackage{amsmath}
\usepackage{amssymb}

\usepackage{threeparttable}
\usepackage[ruled,vlined,linesnumbered]{algorithm2e}
\usepackage[varqu,varl]{inconsolata}
\usepackage{pifont}

\usepackage{multirow}
\newcommand{\tool}{\textsc{Praxis}\xspace}

\usepackage{amsthm}\newtheorem{definition}{Definition}

\IEEEoverridecommandlockouts

\title{From Signals to Behaviors: Evidence-Based Android Malware Detection}

\author{
\IEEEauthorblockN{Shiwen Song\IEEEauthorrefmark{1},
Yiheng Xiong\IEEEauthorrefmark{1}$^{*}$,
Sen Chen\IEEEauthorrefmark{2},
Xiaofei Xie\IEEEauthorrefmark{1}}
\IEEEauthorblockA{\IEEEauthorrefmark{1}Singapore Management University, Singapore\\
swsong@smu.edu.sg, yihengx98@gmail.com, xfxie@smu.edu.sg}
\IEEEauthorblockA{\IEEEauthorrefmark{2}Nankai University, China\\
senchen@nankai.edu.cn}
\thanks{$^{*}$Corresponding author.}
}

\begin{document}

\maketitle

\begin{abstract}

Android malware remains a persistent threat, and detecting it accurately is a long-standing open problem. Whether an app is malicious depends on what it actually does and the context in which it does it, not on the surface signals it happens to exhibit.
Existing detectors instead reason about proxies for behavior, such as learned features or local code slices, and flag whatever deviates from these proxies as malicious. But deviation is not maliciousness: benign apps that merely look unusual are over-flagged, evolving malware that looks ordinary slips through. 
We argue that detection should be behavior-oriented: recover an app's potentially malicious behaviors and judge which are truly malicious. 
To realize this, we present \tool, which structures detection as a hypothesize--confirm--judge pipeline: it hypothesizes candidate behaviors from coarse static signals, confirms each by grounding it in code evidence verified with program analysis, and judges the confirmed behaviors in context: the user's awareness, the app's functional context, and how they compose into an attack. For a malicious app, \tool returns a verdict and the supported behaviors.
We evaluate \tool against seven baselines across three challenging settings. It achieves the best overall detection performance (87.4\% F1), outperforming the baselines by 18.6--34.8 percentage points. On high-permission benign apps, it reduces the false-positive rate to 13.0\%, a reduction of 41.1--67.0 percentage points compared with the baselines. Beyond binary detection, \tool recovers fine-grained malicious behaviors at 87.3\% F1, outperforming prior behavior-level approaches by 56.5--73.4 percentage points. Ablation studies show that each stage of the pipeline contributes to the final performance.

\end{abstract}

\section{Introduction}

Android runs on the majority of the world's mobile devices and mediates much of users' financial and personal activity, which makes it the prime target of malicious software. Despite app-store vetting and on-device protection, Android malware remains a persistent threat: more than 2.67 million mobile attacks were blocked in the first quarter of 2026 alone~\cite{malwarevolumn}, and the apps that slip past these defenses inflict real privacy and financial harm. Detecting them accurately is therefore a long-standing and still-open security problem.

At its core, what makes an app malicious is not the signals it exhibits, such as its permissions, API calls, or code features, but \emph{what the app does and the context in which it does it}. 
The same operation can be legitimate or malicious depending on the situation in which it runs: sending an SMS message is benign when the user knowingly triggers it, but malicious when an app silently sends premium-rate messages behind the user's back; and even an operation the user never sees can be legitimate, as when a cloud-backup app silently uploads files in the background in line with its advertised function, while that same silent upload is spyware in an app that offers no such function.
Maliciousness is thus a property of a \emph{behavior judged in context}: whether the user is aware the behavior runs, and whether it falls within the app's functional context. An effective detector should reason at this level: identify the behaviors an app actually performs, and judge each against its context.

By this standard, existing detectors fall short because they reason about \emph{proxies} for behavior rather than behavior itself. 
\emph{Learning-based} detectors classify an app from features such as permissions, sensitive-API usage, or API-call graphs~\cite{arp2014drebin,wu2019malscan,he2022msdroid,zheng2024maskdroid}; these features only \emph{approximate} behavior, so feature-rich benign apps that legitimately request broad permissions are over-flagged~\cite{onwuzurike2019mamadroid}, while malware that reimplements its logic around the monitored features slips through, and the model emits an opaque label that drifts as malware evolves~\cite{pendlebury2019tesseract} and explains nothing. Even approaches that explicitly consider context, such as AppContext~\cite{yang2015appcontext}, encode it as hand-crafted features for a classifier rather than judging recovered behaviors, and so inherit the same proxy limitation.

More recent \emph{LLM-based} methods~\cite{qian2025lamd,li2025foredroid} begin to recover behavioral content and natural-language explanations: LAMD~\cite{qian2025lamd} summarizes backward slices taken around sensitive APIs, and ForeDroid~\cite{li2025foredroid} flags sensitive-API call chains that deviate from a benign reference distribution. Yet slices and call chains are only partial, local views: neither assembles them into a complete, code-grounded behavior, and neither judges that behavior in the app's actual context. Across all three categories, then, no detector recovers \emph{what an app does}, the very basis on which maliciousness is defined.

We therefore argue that detection should be \emph{behavior-oriented}: instead of recognizing how an app \emph{looks}, reconstruct what it \emph{does} and judge those behaviors in context, much as a human analyst would. This is compelling for three reasons. It is \emph{explainable}: each verdict is backed by concrete behaviors and the code evidence that realizes them, not an opaque score. It is \emph{faithful to how maliciousness is actually defined}: it judges behavior against user awareness and functional context, the same criteria an analyst applies. And it is \emph{robust}: a malicious app can drop conspicuous permissions, obfuscate its APIs, and mimic a benign one~\cite{he2023efficient, li2023black, song2025fcghunter}, but it cannot drop the harmful behavior itself insofar as that behavior is realized in its code, so a behavior-level detector resists the feature drift and evasion that erode signal-based detectors.

Realizing behavior-oriented detection, however, is hard. In principle it is straightforward (enumerate every behavior an app performs and judge each in its context), but carrying this out is infeasible, for three reasons. \emph{(C1) Malicious behavior is sparse in a large behavior space.} A single app performs a wide range of functionalities, the overwhelming majority of them benign, while malicious behavior, when present, is only a sparse few buried among them~\cite{samhi2022difuzer}. Recovering all the behaviors an app performs and then pinpointing the malicious ones is therefore hard: the space to search is large, and the malicious target within it is rare~\cite{wu2021homdroid}. 
\emph{(C2) There is a gap between semantic behaviors and code evidence.} 
A behavior describes what an app does at the semantic level, but proving that it is actually implemented requires concrete code evidence. Bridging this gap is difficult because a behavior rarely maps to a single code location; instead, its implementation may be scattered across multiple functions and execution paths.
\emph{(C3) A behavior's context is hard to recover.} Even once a behavior is established, judging it requires \emph{context} the app never states (whether the user is aware the behavior runs and whether it fits the app's functional context) and, beyond any single behavior, how several behaviors combine into a coherent attack. Recovering this implicit context for each behavior, and reasoning over the behaviors jointly, is itself difficult.

To address these challenges, we introduce \tool, a behavior-oriented Android malware detector built around a \emph{hypothesize--confirm--judge} pipeline.
\tool first \emph{hypothesizes}: it extracts heterogeneous coarse signals (manifest permissions and components, intent actions and sensitive-API call sites from code, and native function names) and uses an LLM as an open-world generator to propose a bounded set of candidate behaviors. It then \emph{confirms} each candidate by locating the functions that could implement it, and using program analysis together with the LLM to verify that they connect into an evidence chain that actually realizes the behavior, discarding any candidate it cannot ground. Finally, it \emph{judges} each confirmed behavior against the user's awareness of its trigger and the app's functional context, flagging the app only when the surviving behaviors compose into a coherent attack pattern. 
\tool outputs a malicious/benign verdict and, for a malicious app, the behavior records behind it.

We evaluate \tool against six learning- and LLM-based baselines across three deployment-stress settings. \tool achieves the best overall detection performance with 87.4\% F1, outperforming the baselines by 18.6--34.8 percentage points, while reducing the false-positive rate on privileged benign apps by 41.1--67.0 percentage points. For fine-grained malicious behavior recovery, \tool achieves 87.3\% F1, outperforming ProMal and ForeDroid by 56.5 and 73.4 percentage points, respectively. Ablation and cross-model studies further show that each stage of \tool contributes substantially to its performance and that the framework remains effective across different LLMs.

In summary, this paper makes the following contributions:
\begin{itemize}[leftmargin=1.2em,itemsep=2pt,topsep=2pt]
  \item We cast Android malware detection as \emph{behavior-oriented, context-aware analysis}: establishing maliciousness by recovering the behaviors an app performs and judging them in context, rather than classifying surface signals, improving both detection accuracy and explainability.
  \item We present \tool, a hypothesize--confirm--judge pipeline that proposes candidate behaviors with an LLM, keeps only those that program analysis can ground in code as \emph{entry $\rightarrow$ source $\rightarrow$ effect} evidence chains, and judges the grounded behaviors against user awareness, functional context, and attack composition.
  \item We show that \tool achieves state-of-the-art malware detection across three deployment-stress settings and identifies fine-grained malicious behaviors beyond the reach of prior detectors, with an ablation isolating the contribution of each stage.
\end{itemize}

\section{Background}

\subsection{Problem Definition}
\label{sec:problem-definition}

Our goal is twofold: to decide whether an APK is malicious and, if so, to recover the behaviors behind that decision.
A \emph{behavior} is a security-relevant action an app performs, described semantically, what the app does, such as ``capture incoming SMS messages and forward them to a remote server.'' A behavior is a semantic notion, independent of any particular implementation.

To decide whether a behavior is actually present, and to make that decision auditable, the behavior must be grounded in code. We call the code that realizes a behavior its \emph{evidence}: the functions that carry the behavior out and the execution paths connecting them. A behavior with no such evidence in an APK is not performed by it.


A behavior, even when grounded, is not malicious or benign on its own; that depends on its \emph{context}, the circumstances under which it runs. We consider two aspects: whether the user is \emph{aware} the behavior runs, that is, whether it is triggered and disclosed through the interface rather than executed silently in the background; and whether the behavior is consistent with the app's \emph{functional context}, the functionality it openly offers. The same behavior flips between benign and malicious as this context changes.



We bundle these into a \emph{behavior record} $u=\langle b, \pi, c\rangle$: a behavior $b$, its code evidence $\pi$, and its context $c$. A malicious app's behaviors rarely act alone; they compose into an overall \emph{attack pattern} that the detector should surface together with the behaviors realizing it. Malware detection is then to recover an APK's behavior records and judge them jointly.



\begin{definition}
\label{def:problem}
Given an APK $a$, recover its grounded behavior records
$\mathcal{R}(a)=\{u_1,\ldots,u_n\}$, with each $u_i=\langle b_i,\pi_i,c_i\rangle$,
and apply a judgment function
\[ M(\mathcal{R}(a)) = \langle Y, P, \mathcal{R}^{+}\rangle, \]
where $Y \in \{\textit{malicious}, \textit{benign}\}$ is the APK-level
verdict, $P$ is the \emph{attack pattern} the app realizes (defined only
when $Y=\textit{malicious}$), and $\mathcal{R}^{+}\subseteq\mathcal{R}(a)$
is the subset of behavior records that compose into $P$.
\end{definition}

\subsection{Motivating Example}
\label{sec:motivating-example}

\begin{figure}[t]
  \centering
  \includegraphics[width=\linewidth]{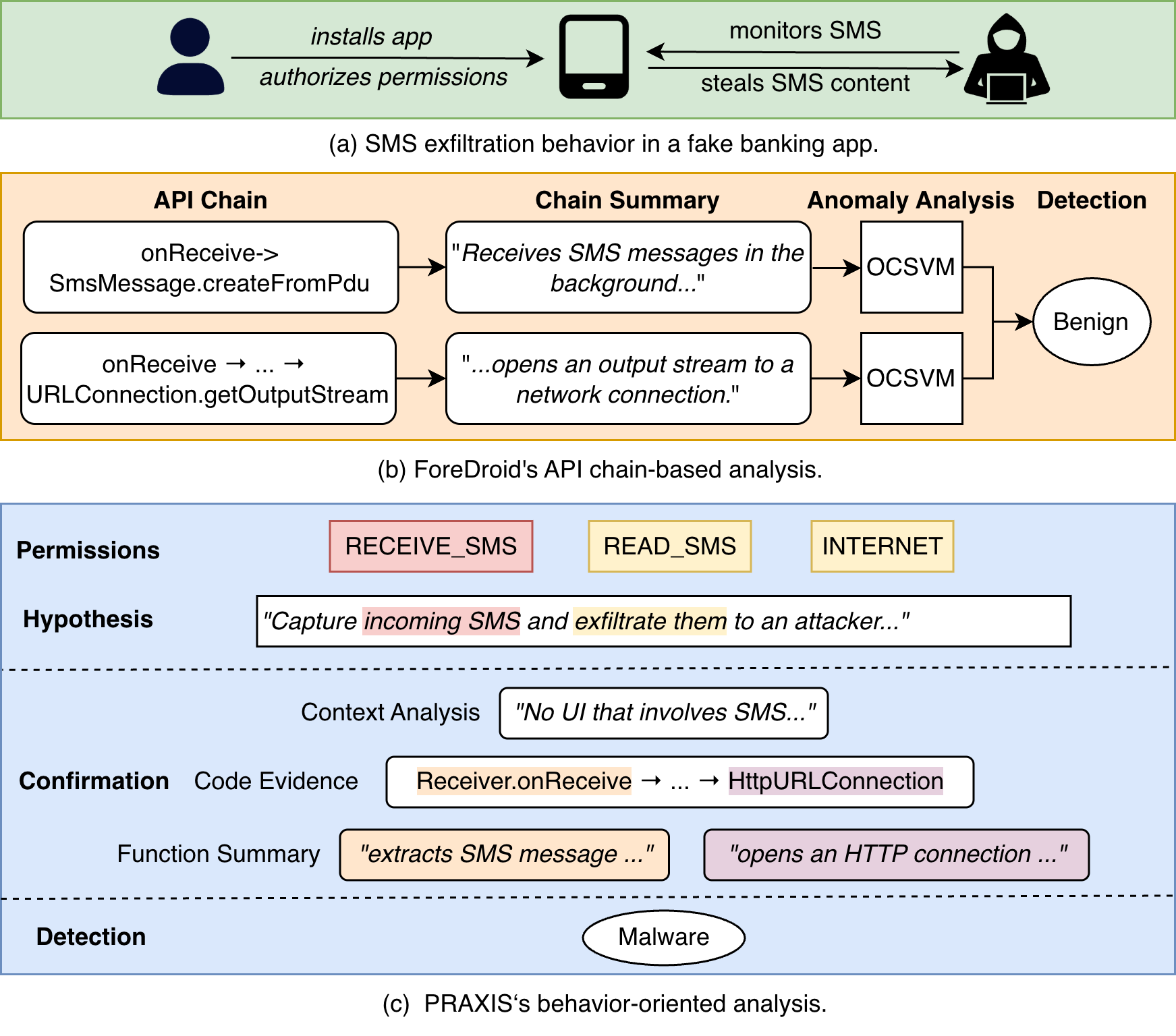}
  \caption{Our motivating example.}
  \label{fig:motivating}
\end{figure}

Figure~\ref{fig:motivating} (a) shows an Android malware sample~\cite{irata_muha2xmad} disguised as a mobile banking app. It presents a fake login screen so the user takes it for a genuine bank. Once installed and granted its permissions, however, it silently monitors incoming SMS in the background, extracts their contents, and forwards them over HTTP to a remote server, carrying out SMS exfiltration. What makes this malicious is not the SMS access itself (a messaging app reads SMS too) but that the access runs \emph{without the user's awareness} and \emph{outside the app's banking purpose}, and that reading and exfiltration \emph{compose} into a single attack: precisely the behavior-judged-in-context on which maliciousness turns.

\noindent\textbf{Limitations of state-of-the-art approaches.}
Both learning-based and LLM-based detectors~\cite{arp2014drebin, wu2019malscan, he2022msdroid, zheng2024maskdroid, qian2025lamd, li2025foredroid} miss this malware, for reasons that differ by approach. 

First, the learning-based detectors (Drebin, MalScan, MsDroid, MaskDroid) do not reason about what the app does. They learn a feature distribution from past malware and judge an app by how its features fit that distribution, taking a statistical pattern as a stand-in for malice. Yet a feature pattern only correlates with maliciousness, and concept drift breaks that correlation as malware evolves: this sample's SMS- and network-related features are each common in benign apps, and their malicious combination is under-represented in the training data, so its features look ordinary and the detectors clear it. 

Second, the LLM-based detectors (LAMD, ForeDroid) attempt to reason about the app's behavior, but still fail to recover this attack behavior.
LAMD summarizes the app bottom-up, one sensitive API at a time, then judges the whole APK from these summaries. Each sensitive API is reached through many call paths, so its per-API summary overlooks the single malicious path and reports a benign intent. The final judgment then weighs all these summaries together, where too many individually benign APIs bury the attack.
ForeDroid breaks the app into entry-to-sensitive-API chains and judges each on its own, keeping only the chains whose embedding departs from a benign distribution and discarding the rest as benign (Figure~\ref{fig:motivating}(b)). Here the SMS-reception and HTTP-transmission chains each look benign in isolation, so ForeDroid discards both and never sees that together they form SMS exfiltration.
In short, none of them recovers the complete behavior the app implements, which is what makes it malicious.

\noindent\textbf{Our approach.}
\tool understands malware at the level of behaviors, what the app actually does. It starts from a candidate behavior the app might perform, grounds that behavior in the code that realizes it, and judges whether it is malicious in context. Figure~\ref{fig:motivating}(c) traces this on the sample. From the APK's signals, \tool hypothesizes the behavior ``capture incoming SMS and exfiltrate them to an attacker's server'' and confirms it in the implementation as a reachable \mbox{entry$\rightarrow$source$\rightarrow$effect} chain: the \texttt{SMS\_RECEIVED} receiver, the PDU parsing that reads the message, and the \texttt{HttpURLConnection} write that sends it out.
It then judges this grounded behavior in context: it is triggered by a background broadcast, so it runs \emph{without the user's awareness}, and it falls \emph{outside the banking functionality} the app advertises; the SMS read and the exfiltration thus \emph{compose} into an SMS-exfiltration attack. \tool labels the APK malicious and outputs behavior records behind it---an account an analyst can audit, which the proxy- and fragment-based detectors cannot provide.

\section{Methodology}
\label{sec:method}


\begin{figure*}[t]
  \centering
  \includegraphics[width=\linewidth]{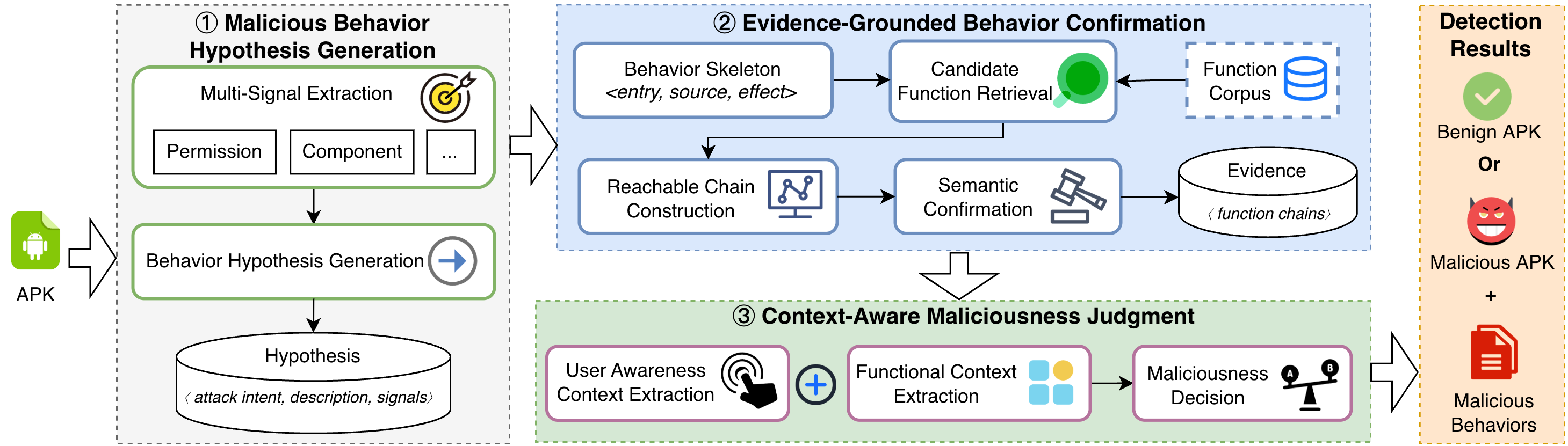}
  \caption{Overview of \tool.}
  \label{fig:overview}
\end{figure*}


Figure~\ref{fig:overview} shows the workflow of \tool, which detects malware in three stages: it \emph{hypothesizes} the malicious behaviors an app may perform, \emph{confirms} each against the code, and \emph{judges} the confirmed behaviors in context, answering challenges C1--C3 in turn.
To address C1, rather than enumerate every behavior in the code, \tool \emph{hypothesizes} (\S\ref{sec:method:hypothesis}): it starts from \emph{signals} that malicious behavior commonly leaves behind and lets an LLM, drawing on its knowledge of Android and malware, propose the behaviors these signals suggest. 
To address C2, \tool \emph{confirms} each hypothesis by grounding it in code (\S\ref{sec:method:evidence}). Following prior Android behavior modeling~\cite{arzt2014flowdroid,yang2015appcontext,liu2022promal}, it casts the behavior as an \emph{entry$\rightarrow$source$\rightarrow$effect} skeleton (the function that initiates the behavior, the one that accesses or operates on the source, and the one that carries out the effect), searches the app's functions for each role, and looks for a reachable chain through the three roles whose code realizes the behavior. The behavior is kept only if such a chain exists, and that chain is its code \emph{evidence}.
To address C3, \tool \emph{judges} the confirmed behaviors in the context (\S\ref{sec:method:judgment}): for each behavior it recovers two kinds of context (whether the user is aware the behavior runs and whether it fits the app's functional context), and then assesses the behaviors together with their evidence and context, deciding whether the app is malicious and, if so, returning the behaviors behind that verdict.

\subsection{Malicious Behavior Hypothesis Generation}
\label{sec:method:hypothesis}

Given an APK, \tool generates the behavior hypothesis that the rest of the pipeline confirms and judges. Because the space of possible behaviors is large and a malicious one may surface anywhere in the code~\cite{li2023black, he2023efficient, song2025fcghunter}, \tool infers them from broad evidence: it reads complementary static signals from the APK and uses an LLM, with its knowledge of Android APIs and malware tactics, to propose the behaviors these signals jointly suggest. Each proposal is a \emph{hypothesis}, paired with the signals that support it, that the next stage then confirms against the code.

\begin{figure}[t]
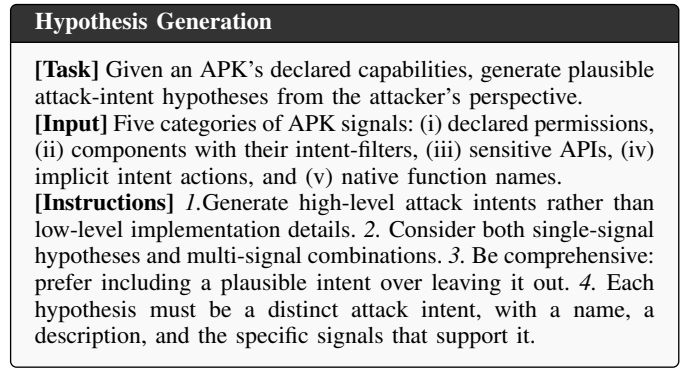

\centering
\begin{tcolorbox}[
    colback=gray!5,
    colframe=black,
    coltitle=white,
    colbacktitle=black!80,
    fonttitle=\bfseries\small,
    title=Hypothesis Generation,
    boxrule=0.5pt,
    arc=2pt,
    left=6pt,right=6pt,top=4pt,bottom=4pt,
    width=\linewidth,
    fontupper=\small
]
\textbf{[Task]} Given an APK's declared capabilities, generate plausible attack-intent hypotheses from the attacker's perspective.

\textbf{[Input]} Five categories of APK signals: (i) declared permissions, (ii) components with their intent-filters, (iii) sensitive APIs, (iv) implicit intent actions, and (v) native function names.

\textbf{[Instructions]} \textit{1.}Generate high-level attack intents rather than low-level implementation details. \textit{2.} Consider both single-signal hypotheses and multi-signal combinations. \textit{3.} Be comprehensive: prefer including a plausible intent over leaving it out. \textit{4.} Each hypothesis must be a distinct attack intent, with a name, a description, and the specific signals that support it.
\end{tcolorbox}
\caption{The prompt for behavior hypothesis generation.}
\label{fig:prompt-hypothesis}
\end{figure}

Specifically, \tool extracts five signals from three sources, shown in Figure~\ref{fig:prompt-hypothesis}. From the \texttt{AndroidManifest.xml}, it takes the requested \emph{permissions}, the capabilities the APK may use, and the declared \emph{components}, the points where it can be entered. From the bytecode, which it analyzes in Soot's IR~\cite{vallee2010soot} because many behaviors are never declared in the manifest, it takes \emph{implicit intent actions}, the APK's interactions with system services, and \emph{sensitive API calls}, identified with the MalScan list~\cite{wu2019malscan}, the security-relevant operations it performs. From the native libraries, it uses Ghidra~\cite{ghidra} to extract \emph{native function names}, since malware often hides security-relevant logic in native code~\cite{samhi2022jucify, xi2024gnndroid,ruggia2025dark}.
 Figure~\ref{fig:module1}(a) illustrates these signals on the motivating example.

\tool prompts the LLM to read these signals from an attacker's perspective and name the attack intents they suggest (Figure~\ref{fig:prompt-hypothesis}). It asks for high-level goals, such as SMS interception, leaving the implementing techniques to the later stages. The prompt balances coverage against speculation: it pushes the LLM to propose intents broadly, including non-obvious ones, while requiring each to be backed by one or more of the observed signals. The signals only seed this proposal: although some are drawn from fixed lists, such as the MalScan sensitive-API set, the LLM may hypothesize behaviors beyond any predefined catalog, so coverage is not bounded by a fixed rule set as in catalog-based detectors. The output is a set of hypotheses, each an attack intent with a short description and its supporting signals. For example, Figure~\ref{fig:module1}(b) shows one hypothesis generated for 
the motivating example, ``SMS Theft for OTP Interception'': the
\texttt{RECEIVE\_SMS}, \texttt{READ\_SMS}, and \texttt{INTERNET} permissions together grant the capability to capture and 
exfiltrate messages, the exported broadcast receiver listening on \texttt{SMS\_RECEIVED} provides an entry point that fires automatically on every message arrival, and the sensitive API \texttt{SmsMessage.createFromPdu}, reachable from this receiver, 
enables PDU parsing of the captured messages.

\begin{figure}[t]
\centering
\captionsetup[subfigure]{skip=2pt}   

\begin{subfigure}{\linewidth}
\begin{tcolorbox}[
    colback=gray!5,
    colframe=black,
    boxrule=0.5pt,
    arc=2pt,
    left=6pt,right=6pt,top=4pt,bottom=4pt,
    width=\linewidth,
    fontupper=\small\raggedright
]
\textbf{[Permissions]} \texttt{RECEIVE\_SMS}, \texttt{READ\_SMS}, \texttt{INTERNET}, \ldots

\textbf{[Components]} \texttt{io.flutter.plugins.Receiver} (\texttt{BroadcastReceiver}, exported), \ldots

\textbf{[Implicit Intent Actions]} \texttt{io.flutter.plugins.Receiver} $\to$ \texttt{android.provider.Telephony.SMS\_RECEIVED}, \ldots

\textbf{[Sensitive APIs]} \texttt{android.telephony.SmsMessage.createFromPdu}, \texttt{android.provider.Settings\$Secure.getString}, \ldots

\textbf{[Native function names]} (none)
\end{tcolorbox}
\caption{Signals extracted from the APK.}
\end{subfigure}

\vspace{4pt}

\begin{subfigure}{\linewidth}
\begin{tcolorbox}[
    colback=gray!5,
    colframe=black,
    boxrule=0.5pt,
    arc=2pt,
    left=6pt,right=6pt,top=4pt,bottom=4pt,
    width=\linewidth,
    fontupper=\small\raggedright
]
\textbf{Attack Intent:} SMS Theft for OTP Interception

\textbf{Description:} Capture incoming SMS (especially bank/2FA OTPs) and exfiltrate them to an attacker to bypass authentication or confirm fraudulent transactions. The app can directly receive SMS broadcasts and read message contents.

\textbf{Supporting Signals:}
\begin{enumerate}[leftmargin=18pt,itemsep=2pt,topsep=2pt,label=\arabic*.]
    \item Permissions: \texttt{RECEIVE\_SMS}, \texttt{READ\_SMS}, \texttt{INTERNET}
    \item Exported receiver listening to \texttt{android.provider.Telephony.SMS\_RECEIVED} (\texttt{io.flutter.plugins.Receiver})
    \item Sensitive API \texttt{SmsMessage.createFromPdu} reachable from the SMS broadcast receiver enables PDU parsing of incoming messages
\end{enumerate}
\end{tcolorbox}
\caption{Hypothesis generated by the LLM.}
\end{subfigure}

\caption{A hypothesis from the motivating example.}
\label{fig:module1}
\end{figure}

\subsection{Evidence-grounded Behavior Confirmation}
\label{sec:method:evidence}

A hypothesis is only a natural-language guess; this stage confirms it by finding the code that realizes it, and keeps the behavior only if such code exists. This is challenging for two reasons. First, the hypothesis is only a natural-language description, with no pointer to the code that realizes it: most of its signals give no code anchor, since a requested permission or declared component names a capability without showing where the app exercises it. Second, even searching the code with the behavior hypothesis is too coarse, because the hypothesis describes a complete behavior, whereas the code implements it through multiple functions, each corresponding to only one semantic role.

\tool recovers it guided by an \emph{entry$\rightarrow$source$\rightarrow$effect} skeleton of the behavior: the \emph{entry} that triggers it, the \emph{source} of sensitive data or state it acts on, and the \emph{effect} it produces. We call the entry, source, and effect the three \emph{roles} of the skeleton: the parts the implementing code must fill, in that order. 
In Algorithm~\ref{alg:evidence}, \tool first derives this skeleton from the hypothesis with the LLM (\textit{Skeleton}, Line~1): from the attack intent, description, and supporting signals, it specifies for each role what the implementing code should do, anchoring a role in the hypothesis's signals wherever they pin it down. In Figure~\ref{fig:module1}(b), for instance, the entry is the callback of the exported \texttt{SMS\_RECEIVED} receiver, the source parses the incoming message via \texttt{SmsMessage.createFromPdu}, and the effect sends the message out over the network. Guided by this skeleton, \tool (i) searches the app's functions for candidates for each role (\textit{Retrieve}, Line~2; \S\ref{sec:method:retrieval}); (ii) keeps the candidate triples whose entry, source, and effect are wired into a reachable \emph{entry$\rightarrow$source$\rightarrow$effect} chain (\textit{Reach}, Lines~3--5; \S\ref{sec:method:structure_verification}); and (iii) asks an LLM to confirm from their code which chains realize the behavior, keeping the confirmed chains as the behavior's evidence $\pi$ and assigning it a support label $\ell$ (\textit{SemanticConfirmation}, Line~7; \S\ref{sec:method:semantic_verification}). A hypothesis whose roles admit no reachable chain is discarded (Line~6); the rest pass on with their evidence. 

\begin{algorithm}[t]
\DontPrintSemicolon
\SetKwInOut{KwIn}{Input}
\SetKwInOut{KwOut}{Output}
\caption{Evidence-grounded behavior confirmation}
\label{alg:evidence}
\KwIn{Hypothesis $h = (\textit{intent}, \textit{description}, \textit{signals})$;
function corpus $\mathcal{C}$.}
\KwOut{Evidence $\pi$ and support label $\ell$.}

$(e, s, f) \leftarrow \textit{Skeleton}(h)$\;
$F_e \leftarrow \textit{Retrieve}(e, \mathcal{C})$;\quad
$F_s \leftarrow \textit{Retrieve}(s, \mathcal{C})$;\quad
$F_f \leftarrow \textit{Retrieve}(f, \mathcal{C})$\;
\BlankLine

$T \leftarrow \emptyset$\;
\ForEach{$(f_e, f_s, f_f) \in F_e \times F_s \times F_f$}{
  \lIf{$\textit{Reach}(f_e, f_s) \wedge \textit{Reach}(f_s, f_f)$}{$T \leftarrow T \cup \{(f_e, f_s, f_f)\}$}
}
\lIf{$T = \emptyset$}{\Return $\bot$}
\BlankLine

$(\pi, \ell) \leftarrow \textit{SemanticConfirmation}(h, T)$\;

\Return $(\pi, \ell)$\;



\end{algorithm}

\subsubsection{Candidate Function Retrieval}
\label{sec:method:retrieval}

\label{sec:method:corpus}
Given a hypothesis, this step retrieves the candidate functions that may fill each role of its skeleton, which the later steps then check.
Because the behavior is realized across its entry, source, and effect, \tool searches for the three roles separately, so that none goes unsearched.

For each APK, \tool builds a retrieval corpus that represents every function in two ways.
For each app-specific function, \tool uses an LLM to generate a concise behavioral summary that preserves Android-specific cues (e.g., framework APIs and component types), and stores the summary with the function's decompiled code.
TPL functions, identified by LibRadar~\cite{ma2016libradar}, are kept as decompiled code only, since summarizing the many generic library functions would add cost and retrieval noise.

To search the corpus, \tool prompts an LLM to build two complementary queries for each role, one against each representation. A \emph{semantic query}, derived from the role's specification, describes how a function filling that role would behave and is matched against the summaries. A \emph{lexical query} lists that role's concrete anchors, such as API names, class or method names, and intent actions, and is matched against the decompiled code.
For example, in Figure~\ref{fig:module1}(b), the source role's semantic query reads ``function that parses SMS PDU and
reads message body and sender,'' while its lexical query lists \texttt{SmsMessage.createFromPdu}.




\tool scores each candidate function $f$ for a role $r$ by combining semantic and lexical similarity:
\[
\mathrm{score}(f,r) =
\alpha \cdot \mathrm{sim}_{\mathrm{sem}}(f,r)
+
\beta \cdot \mathrm{sim}_{\mathrm{lex}}(f,r).
\]
Here, $\mathrm{sim}_{\mathrm{sem}}$ is the cosine similarity between SBERT~\cite{reimers2019sentence} embeddings of the role query and the function summary, while $\mathrm{sim}_{\mathrm{lex}}$ is the BM25~\cite{robertson2009probabilistic} score between the lexical anchors and the decompiled code. \tool keeps the top-$k$ functions for each role as its candidate set $F_r$ (\textit{Retrieve} in Algorithm~\ref{alg:evidence}). The weights favor semantic similarity, with lexical anchors serving as complementary cues; $k$ trades coverage against the cost of checking candidate chains. We report $\alpha$, $\beta$, and $k$ in our experimental setup.

\subsubsection{Reachable Chain Construction}
\label{sec:method:structure_verification}

Retrieval ranks candidates for each role by similarity alone, so not every candidate belongs to the behavior. \tool keeps only those that fit together into a chain: the functions realizing one behavior must connect from entry through source to effect, so a triple whose functions cannot reach one another in that order is spurious. \tool therefore keeps the candidate triples whose functions form a reachable chain,
\[
\begin{aligned}
T = \bigl\{\, (f_e, f_s, f_f) &\in F_e \times F_s \times F_f \bigm| \\
&\textit{Reach}(f_e, f_s) \wedge \textit{Reach}(f_s, f_f) \,\bigr\},
\end{aligned}
\]
where $\textit{Reach}(f, f')$ holds when $f'$ is reachable from $f$ (\textit{Reach} in Algorithm~\ref{alg:evidence}).

Because Android control and data flow propagate through more than direct calls, \tool decides reachability through three channels:
\[
\begin{aligned}
\textit{Reach}(f, f') \Longleftrightarrow{} &R_{\mathrm{call}}(f, f') \vee R_{\mathrm{icc}}(f, f') \\
&\vee\, R_{\mathrm{state}}(f, f') \vee f = f'.
\end{aligned}
\]
\emph{Call reachability} $R_{\mathrm{call}}$ holds when one function reaches the other through call edges.
\emph{ICC reachability} $R_{\mathrm{icc}}$ holds when the two lie in different components joined by an inter-component communication transition~\cite{li2015iccta}.
\emph{State reachability} $R_{\mathrm{state}}$ holds when one writes and the other reads a shared object, such as a class field, a \texttt{SharedPreferences} key, or a \texttt{ContentProvider} URI; we check these high-confidence shared objects directly rather than running def-use analysis over the whole APK, which reduces cost and noise.
The final disjunct $f = f'$ covers a single function that fills both roles.
A hypothesis with no reachable chain ($T = \emptyset$) is discarded; the rest pass to semantic confirmation.

\subsubsection{Semantic Confirmation}
\label{sec:method:semantic_verification}



A connection by a program relation does not guarantee semantic relevance: two functions can be wired together yet have nothing to do with the hypothesized behavior. \tool therefore confirms each candidate chain in two steps. First, for each adjacent pair on the chain, it reads the two endpoints' decompiled code and asks whether they carry out that step of the behavior, dropping any pair whose implementation does not hold up. Second, it asks whether the surviving chain realizes the behavior as a whole, exhibiting each part, such as the data accessed and the APIs invoked, and connecting them into one coherent realization. \tool then labels each chain \emph{supported}, \emph{partial}, or \emph{not-supported}, discards the \emph{not-supported} ones, and carries the rest, with their label, to the maliciousness judgment.

\subsection{Context-Aware Maliciousness Judgment}
\label{sec:method:judgment}

This stage judges the confirmed behaviors in the context the APK leaves implicit, deciding whether it is malicious and identifying the behaviors behind that decision. A confirmed behavior says only what the APK does, not whether it is malicious: as in the motivating example (Fig.~\ref{fig:motivating}), reading incoming SMS is benign in a messaging app the user opens, but malicious when an app does it silently and forwards the messages to a remote server. Moreover, one behavior alone rarely settles maliciousness; malware is marked by several behaviors combining into an attack~\cite{christodorescu2007mining, fredrikson2010synthesizing}.
\tool therefore recovers, for each behavior, two kinds of context the APK leaves implicit: whether the user is aware the behavior runs (\S\ref{sec:method:ui}) and whether it fits the app's functional context (\S\ref{sec:method:business}). It then judges the behaviors together with their evidence and context (\S\ref{sec:method:global}). Rather than discard a behavior the moment it looks justified, \tool attaches its context and weighs all behaviors jointly, so that a behavior innocuous on its own is still considered as part of a possible attack.

\subsubsection{User-Awareness Context Extraction}
\label{sec:method:ui}

Malware works by hiding its harmful behaviors from the user, so whether the user is aware a behavior runs is a strong contextual cue; \tool recovers this awareness as context for the judgment.
\tool first recovers how the behavior is triggered, tracing backward from its evidence $\pi$ to the Android entry points in the enhanced call graph~\cite{yan2022iccbot}.
A behavior reached only through background events, such as a system broadcast, a scheduled task, or a long-running service, starts without the user and is labeled \emph{non-aware}. A behavior whose entry points cannot be recovered is labeled \emph{unknown}. A behavior the user starts through the UI is examined further to see whether the interface discloses what it does.

For such a UI-triggered behavior, \tool builds a semantic context for the triggering UI: it summarizes the Activity with an LLM to capture the screen's purpose and, for a widget-triggered behavior, adds the widget's visible text. This combines screen-level and widget-level cues, a finer basis than the widget labels or isolated UI strings used in prior work~\cite{xi2019deepintent, li2025foredroid, li2025ui}. \tool then determines whether this context discloses the behavior, that is, whether the behavior falls within what the UI leads the user to expect; if so, the behavior is labeled \emph{aware}, otherwise \emph{non-aware}.
Each behavior thus carries an awareness label (\emph{aware}, \emph{non-aware}, or \emph{unknown}) into the final judgment, rather than being filtered out here.

\subsubsection{Functional Context Extraction}
\label{sec:method:business}

\tool also recovers whether each behavior is consistent with the app's \emph{functional context}, what the app openly does for its users, as a second kind of context.
It infers this context chiefly from the screen-level UI summaries of the app's user-visible Activities, supplemented by the app's label, store category, and main-package class names.
Whether a behavior fits this context, and how far it departs from it, is attached to the behavior as functional context for the final judgment (\S\ref{sec:method:global}).

\begin{figure}[t]
\centering
\begin{tcolorbox}[
    colback=gray!5,
    colframe=black,
    coltitle=white,
    colbacktitle=black!80,
    fonttitle=\bfseries\small,
    title=Maliciousness Decision,
    boxrule=0.5pt,
    arc=2pt,
    left=6pt,right=6pt,top=4pt,bottom=4pt,
    width=\linewidth,
    fontupper=\small
]
\textbf{[Task]} Given the risky behaviors, decide whether they compose a recognized Android malware attack pattern.

\textbf{[Input]} For each risky behavior: the evidence with per-function summaries, the trigger context, the attack intent, and the description. 

\textbf{[Instructions]}
\textit{1. Integration over tallying:} reason about how the behaviors compose into a coherent attack lifecycle, not by counting verdicts.
\textit{2. Key behavior over completeness:} report the risky behaviors that best characterize the resulting attack pattern, rather than listing every recovered behavior.
\textit{3. Evidence over hypothesis:} when the recovered evidence narrows, refines, or contradicts the intent or description, anchor the characterization in the evidence rather than the original hypothesis.

\end{tcolorbox}
\caption{The prompt for maliciousness decision.}
\label{fig:prompt-judgment}
\end{figure}

\subsubsection{Maliciousness Decision}
\label{sec:method:global}

The two contexts above bear on each behavior in isolation; what remains is how the behaviors relate to \emph{one another}. Reading incoming SMS, for instance, could be legitimate, but once the app also uploads those messages to a remote server, the two together are a clear SMS-exfiltration attack. \tool therefore makes the final decision over all behaviors jointly.

\tool gives the LLM all the confirmed behaviors at once, each with its evidence $\pi$, function summaries, awareness and functional context (\S\ref{sec:method:ui}, \S\ref{sec:method:business}), and hypothesized intent. \tool then prompts it (Figure~\ref{fig:prompt-judgment}) to judge each behavior in its context and to reason how the behaviors compose into an attack, identifying the attack's key behaviors. \tool labels the APK malicious if such an attack emerges, and benign otherwise. For a malicious verdict, it returns the attack pattern and these key behaviors, each with its evidence and a description matched to its code, as a code-grounded account of the decision. Figure~\ref{fig:module3-output} illustrates this on the motivating example (\S\ref{sec:motivating-example}): five behaviors composing into an SMS-exfiltration attack pattern.



\begin{figure}[t]
\centering
\begin{tcolorbox}[
    colback=gray!5,
    colframe=black,
    boxrule=0.5pt,
    arc=2pt,
    left=6pt,right=6pt,top=4pt,bottom=4pt,
    width=\linewidth,
    fontupper=\small\raggedright
]
\textbf{Attack Pattern:} SMS-based OTP Interception and Exfiltration

\textbf{Key Behaviors:}
\begin{enumerate}[leftmargin=18pt,itemsep=2pt,topsep=2pt,label=\arabic*.]
    \item Silent SMS interception via a background broadcast receiver (\texttt{io.flutter.plugins.Receiver.onReceive}) listening on \texttt{android.provider.Telephony.SMS\_RECEIVED} (intent-filter priority 999), triggered automatically by the Android framework without user interaction.
    \item PDU parsing of incoming messages to extract body, sender, and service-center address (\texttt{SmsMessage.createFromPdu}, \texttt{getMessageBody}, \texttt{getOriginatingAddress}).
    \item Device fingerprinting via Android ID lookup (\texttt{Receiver.c}, \texttt{Settings\$Secure.getString}), packaged alongside the captured SMS payload.
    \item Background exfiltration of the captured data through a Kotlin coroutine and lambda (\texttt{Receiver.b}, \texttt{s3.a.a}, \texttt{Receiver\$a.a}) terminating in an \texttt{HttpURLConnection} write to a remote endpoint, traversed entirely without UI involvement.
    \item Bank Mellat impersonation as a deception layer: app label ``Hamrah Bank'' (Persian for ``Mobile Bank'') and a Flutter-rendered banking-style login interface (card number, password, password recovery), but no real Bank Mellat backend connection or banking transaction logic.
\end{enumerate}
\end{tcolorbox}
\caption{A recognized attack pattern from the motivating example.}
\label{fig:module3-output}
\end{figure}

\vspace{-15pt}
\section{Evaluation}

We evaluate \tool by investigating the following research questions:

\begin{itemize}[leftmargin=1em,itemsep=2pt,topsep=2pt]
    \item \textbf{RQ1}: How effective is \tool at malware detection compared with existing approaches?
\item \textbf{RQ2}: How well does \tool identify malicious behaviors?
\item \textbf{RQ3}: How does each component of \tool contribute to its effectiveness?
\item \textbf{RQ4}: How does the underlying LLM affect the effectiveness and cost of \tool?
\end{itemize}

\subsection{Experimental Setup}
\label{sec:setup}
\subsubsection{Baselines}
\label{sec:baselines}
We use seven baselines in total: six for malware detection and two for malicious-behavior identification, with ForeDroid serving in both roles. 
For malware detection, we compare \tool against six state-of-the-art Android malware detectors spanning two categories.

\emph{\ding{172} Learning-based detectors.}
We include four learning-based detectors~\cite{gao2024comprehensive,liu2026unraveling}, spanning feature-based (Drebin) and graph-based (MalScan, MsDroid, MaskDroid) detection. Drebin~\cite{arp2014drebin} extracts static features from the manifest and code, encodes them as a binary feature vector, and trains an SVM classifier for malware detection. MalScan~\cite{wu2019malscan} constructs function call graphs, represents each APK using centrality-based features of sensitive APIs, and trains a $k$-nearest-neighbor classifier. MsDroid~\cite{he2022msdroid} extracts sensitive API graphs and trains a GNN-based classifier.
MaskDroid~\cite{zheng2024maskdroid} further improves GNN robustness through masked graph representation learning and contrastive learning.

\emph{\ding{173} LLM-based detectors.}
We include two LLM-based detectors, LAMD~\cite{qian2025lamd} and ForeDroid~\cite{li2025foredroid}, both of which use an LLM to analyze app behavior; for ForeDroid, we use the OCSVM model released by its authors.

For malicious-behavior identification, we compare against two baselines that also produce behavior records: ForeDroid, described above, and ProMal~\cite{wu2025beyond}. ProMal is not designed for APK-level malware detection; it assumes the input APK is already malicious and maps API-level evidence to malicious behavior records using an expert-built knowledge graph.

\subsubsection{Dataset}
\label{sec:dataset}
Our evaluation dataset contains 1{,}033 apps in total, split into three sets, each reflecting a setting that detectors encounter in real-world deployment: \emph{time-shift} (600 apps), the most recent malware (2024--2026); \emph{diverse-family} (163 malware), one sample from each of many distinct families; and \emph{privileged-benign} (270 benign apps), each requesting many sensitive permissions.


\emph{Training and validation data.} \tool requires no training, so this data serves only the learning-based baselines, which we retrain on a unified set since their original training data is unavailable or inconsistent across studies. Following standard collection practice~\cite{pendlebury2019tesseract, liu2026unraveling, li2025foredroid, song2025fcghunter}, we collect apps from AndroZoo~\cite{allix2016androzoo} and label them with VirusTotal~\cite{virustotal}. The training set is balanced across 2021--2023 at 10{,}000 apps per year (30{,}000 in total), and the validation set adds a disjoint 2{,}500 apps per year, on which the baselines reach F1 scores of $97.8\%$ (Drebin), $96.7\%$ (MalScan), $91.5\%$ (MsDroid), and $95.1\%$ (MaskDroid), confirming they are properly trained. Note that the three evaluation sets described next are all disjoint from these.



\emph{Time-shift.} This set evaluates whether detectors generalize to recent APKs.
We use samples from 2024--2026, with 100 malware from MalwareBazaar~\cite{malwarebazaar} and 100 benign APKs from AndroZoo per year, totaling 600 APKs.
For learning-based detectors, this setting measures robustness to concept drift as samples move away from the training distribution~\cite{pendlebury2019tesseract}.
For LLM-based detectors, it tests whether they are also affected by changes in APKs over time, despite not relying on task-specific training.

\emph{Diverse-family.} This set evaluates whether a detector remains effective across diverse Android malware families. We draw from GPMalware~\cite{cao2022rotten} and MalwareBazaar~\cite{malwarebazaar}, together spanning 2015--2026; after discarding samples that lack a reliable family label, we obtain 163 malware spanning 163 distinct families.



\emph{Privileged-benign.} Many legitimate apps, such as security, backup, and device-management tools, genuinely require sensitive permissions, which makes them easy to mistake for malware. This set measures how often a detector falsely flags such apps as malicious. We collect benign APKs from AndroZoo spanning 2015--2025, rank them by the number of declared dangerous permissions, and keep the top 270 statically analyzable ones; each declares at least 9 dangerous permissions, and the set spans 49 Google Play categories.








\subsubsection{Environment}
\label{sec:environment}
\tool, LAMD, and ForeDroid all use DeepSeek-V4-Pro, which balances reasoning quality and inference cost, and running every method on the same model keeps the comparison fair. The one exception is \tool's corpus summarization (\S~\ref{sec:method:corpus}), which runs over every app-specific function and uses the cheaper DeepSeek-V4-Flash to keep this high-volume step affordable. We fix the temperature to 0 for all calls to aid reproducibility. Because LLM outputs can still vary at temperature 0, we run each LLM-based method (\tool, LAMD, and ForeDroid) three times and take the majority verdict over the three runs. For retrieval (\S\ref{sec:method:retrieval}), we empirically set the similarity weights to $\alpha=0.7$ and $\beta=0.3$, favoring semantic over lexical similarity, and keep the top $k=10$ candidates for each role, balancing coverage against verification cost.

\subsubsection{Evaluation Method for RQ1}
We compare \tool against the six baselines from \S~\ref{sec:baselines}. Because the three test sets differ in composition, we report the metric each supports: on the balanced \emph{time-shift} set, F1 together with false-negative rate (FNR) and false-positive rate (FPR); on the malware-only \emph{diverse-family} set, FNR (the fraction of malware missed); and on the benign-only \emph{privileged-benign} set, FPR (the fraction of benign apps flagged as malicious).

\subsubsection{Evaluation Method for RQ2}
We evaluate \tool's behavior identification effectiveness from two angles: a comparison with ForeDroid~\cite{li2025foredroid} and Promal~\cite{wu2025beyond}, the state-of-the-art tools that identify malicious behaviors, and a per-stage analysis of \tool's pipeline.

To support these evaluations, we build a behavior-level ground truth on the \emph{diverse-family} dataset, chosen because every sample is accompanied by a high-quality threat report. Following existing practice~\cite{li2025foredroid}, we first refine the 8 payload categories defined in GPMalware~\cite{cao2022rotten} into 50 fine-grained behavior types. We then manually extract the documented malicious behaviors from each threat report and its APK and annotate them, yielding 962 ground-truth behaviors across the 163 samples.

Two authors, each with more than three years of Android malware analysis experience, independently assign both ground-truth behaviors and candidate behaviors to one of the 50 fine-grained behavior types.
Disagreements in type assignment are resolved through discussion.
For each sample, behavior identification is computed by matching the assigned behavior types of the recovered behaviors against those of the ground-truth behaviors.



We apply this protocol in two settings. In the baseline comparison, we compare the final behavior records produced by \tool and ForeDroid on the 81 samples that both methods correctly classify as malicious, covering 565 ground-truth behaviors. In the per-stage analysis, we apply the same protocol to the outputs of every stage on the \emph{diverse-family} dataset. Inter-annotator agreement reaches Cohen's $\kappa=0.88$ and $0.83$ for the two settings, respectively.

\subsubsection{Evaluation Method for RQ3}

To evaluate the contribution of each major stage in \tool, we construct three ablated variants corresponding to the three stages of our pipeline.
\emph{w/o Multi-Signals} replaces multi-signal behavior hypothesis generation (\S\ref{sec:method:hypothesis}) with hypotheses generated solely from sensitive APIs, evaluating the benefit of multi-signals.
\emph{w/o Verification} removes both structural(\S\ref{sec:method:structure_verification}) and semantic verification (\S\ref{sec:method:semantic_verification}), evaluating the importance of grounding behavior hypotheses in code evidence.
\emph{w/o Context} removes user-awareness judgment (\S\ref{sec:method:ui}), functional context judgment (\S\ref{sec:method:business}), and attack-pattern recognition (\S\ref{sec:method:global}).
It instead uses a plain judgment prompt to decide maliciousness directly from the recovered code evidence and behavior hypotheses, without explicit context-aware reasoning.
We evaluate all variants on the same three evaluation datasets as RQ1 and report the corresponding detection metrics.

\subsubsection{Evaluation Method for RQ4}
We run \tool with three LLMs spanning different capability and cost tiers: GPT-5.2, DeepSeek-V4-Pro (our default), and GPT-5.4-mini. Each substitutes only the detection-reasoning model; corpus summarization (\S~\ref{sec:method:corpus}) is held fixed at the lightweight model (\S~\ref{sec:environment}), so the comparison isolates the reasoning model. We report effectiveness with the same per-set metrics as RQ1, and cost as the average dollar cost per app, split into the fixed summarization cost and the model-dependent judgment cost.

\subsection{Results of RQ1: Malware Detection Performance}
\label{sec:eval:rq1}


\begin{table}[!t]
    \centering
    \caption{Malware Detection under Challenging Scenarios.}
    \label{tab:rq1}
    \renewcommand{\arraystretch}{1.4}
    \resizebox{\columnwidth}{!}{
    \begin{tabular}{c|ccc|c|c|c}
    \hline
    \hline
    \textbf{Method}
    & \multicolumn{3}{c|}{\textbf{Time-Shift}}
    & \textbf{Diverse-Family}
    & \textbf{Privileged-Benign}
    & \textbf{Overall} \\
    \cline{2-7}
    & \textbf{FNR} & \textbf{FPR} & \textbf{F1}
    & \textbf{FNR}
    & \textbf{FPR}
    & \textbf{F1} \\
    \hline

    \rowcolor[HTML]{FCE5CD}
    Drebin
    & 27.3\% & 3.3\% & 82.6\%
    & 25.2\%
    & 65.2\%
    & 68.8\% \\

    MalScan
    & 49.3\% & 8.3\% &63.7\%
    & 39.3\%
    & 60.0\%
    & 55.7\% \\

    \rowcolor[HTML]{FCE5CD}
    MsDroid
    & 44.0\% & 31.0\% & 59.9\%
    & 36.8\%
    & 75.2\%
    & 52.6\% \\

    MaskDroid
    & 31.3\% & 16.7\% & 74.1\%
    & 22.7\%
    & 72.6\%
    & 63.8\% \\

    \rowcolor[HTML]{FCE5CD}
    LAMD

    & 50.0\% & 15.0\% & 60.6\%
& 26.9\%
& 80.0\%
& 55.0\% \\

    ForeDroid

        & 56.0\% & 10.7\% & 56.9\%
    & 37.4\%
    & 54.1\%
    & 53.5\% \\

    \hline
    \rowcolor[HTML]{DCDCDC}
    \textbf{Ours}
    & \textbf{11.7\%} & \textbf{3.3\%} & \textbf{92.2\%}
    & \textbf{20.9\%}
    & \textbf{13.0\%}
    & \textbf{87.4\%} \\
    \hline
    \hline
    \end{tabular}
    }
    \end{table}

Table~\ref{tab:rq1} shows that \tool leads on every setting and overall: the highest F1 on time-shift (92.2\%), the lowest FNR on diverse-family (20.9\%), the lowest FPR on privileged-benign (13.0\%), and the best overall F1 (87.4\%). It exceeds the strongest baseline, Drebin (68.8\%), by 27.0\% and the weakest, MsDroid (52.6\%), by 66.2\% in relative overall F1.


\tool's advantage is that it evaluates maliciousness directly, recovering each behavior from code and judging whether it is malicious in context, exactly what maliciousness depends on.
The baselines instead reach maliciousness only through proxies that correlate with it, a learned feature distribution or a fixed list of sensitive APIs, and each setting is a case where one of these proxies decouples from actual malicious behavior.
A learned distribution goes stale as apps evolve, so the learning-based baselines drift to 59.9--82.6\% F1 on time-shift.
A sensitive-API list misses malware that acts through other code, capping the API-keyed baselines at 52.6--63.8\% overall F1, and fires on benign apps that use those APIs legitimately, raising their FPR to 54.1--80.0\% on privileged-benign.

\subsection{Results of RQ2: Malicious Behavior Identification}
\label{sec:eval:rq2}

\subsubsection{Fine-Grained Behavior Identification Performance}
As shown in Table~\ref{tab:behavior_detection_performance}, \tool substantially outperforms both baselines in fine-grained behavior identification, achieving an F1 of 87.3\% compared with 30.8\% for ProMal and 13.9\% for ForeDroid. This improvement is primarily due to a large precision gain (94.8\% vs.\ 19.1\% and 8.5\%), while maintaining comparable recall to ProMal (80.9\% vs.\ 78.9\%) and substantially outperforming ForeDroid (38.5\%). Although ForeDroid and ProMal generate far more candidates (1,590 and 5,664, respectively), most are spurious and do not correspond to genuine malicious behaviors, indicating that recovering low-level API chains or knowledge-graph operations alone is insufficient to reconstruct complete malicious behaviors.

\noindent
\textbf{FP Analysis.}
Most of the 30 false positives come from third-party SDKs bundled in the APK. Such SDKs perform sensitive operations, such as advertising-ID collection, but \tool cannot see their internal implementation and so cannot tell whether an operation is benign library functionality or part of the app's malicious behavior. It therefore retains these sensitive operations as code-grounded evidence, which can push the verdict to malicious. A promising direction for future work is to understand these third-party SDKs more accurately, which would reduce such false positives.

\noindent
\textbf{FN Analysis.}
The 118 false negatives produced by \tool mostly stem from content that is loaded only at runtime, which static analysis cannot see: a page an attacker serves into a WebView, or a payload downloaded or decrypted on the fly. In these cases \tool observes the loading logic, such as the WebView setup or the load call, but not the content or code that actually runs, and it is that runtime content that determines maliciousness.



\begin{table}[!t]
\centering
\caption{Fine-Grained Behavior Identification Comparison. }
\label{tab:behavior_detection_performance}
\renewcommand{\arraystretch}{1.4}
\resizebox{0.9\columnwidth}{!}{ 
\begin{tabular}{c|ccc|ccc}
\hline
\hline
\textbf{Method} & \textbf{\#Detected} & \textbf{\#FP} & \textbf{\#FN} & \textbf{Precision} & \textbf{Recall} & \textbf{F1} \\
\hline

Promal     & 5664 & 4581 & 119 & 19.1\% & 78.9\% & 30.8\% \\

ForeDroid & 1590 & 1473 & 334 & 8.5\% & 38.5\% & 13.9\% \\

\hline
    \rowcolor[HTML]{DCDCDC}
Ours & 501 & 30 & 118 & 94.8\% & 80.9\% & 87.3\% \\

\hline
\hline
\end{tabular}
}
\end{table}

\subsubsection{Per-Stage Behavior Analysis}

Table~\ref{tab:stage_behavior_detection} shows that \tool progressively refines behavior candidates across its three stages. This process sharply improves precision from 46.5\% to 96.8\%, while retaining 79.5\% of the ground-truth behaviors. Overall, the pipeline improves F1 by 24.4 percentage points, showing that high-recall hypothesis generation can be effectively turned into precise behavior recovery through evidence confirmation and judgment.

Manual inspection shows that the remaining recall loss comes from different sources at different stages.
Generation misses 25 behaviors, mainly uncommon or family-specific behaviors whose signal combinations do not trigger explicit hypotheses.
Confirmation removes 108 behaviors, mainly because static analysis cannot recover complete evidence chains, or because retrieval returns code that is semantically related to the hypothesis but does not contain the evidence required to support it.
Judgment removes another 64 behaviors because their malicious intent is weak, ambiguous, or overshadowed by the dominant attack chain.

\begin{table}[!t]
\centering
\caption{Per-Stage Behavior Identification Performance.}
\label{tab:stage_behavior_detection}
\renewcommand{\arraystretch}{1.4}
\resizebox{0.75\columnwidth}{!}{
\begin{tabular}{c|ccc}
\hline\hline
\textbf{Metric}
& \textbf{Generation}
& \textbf{Confirmation}
& \textbf{Judgement} \\
\hline
Precision & 46.5\% & 84.0\% & \textbf{96.8\%} \\
Recall    & \textbf{97.4\%} & 86.2\% & 79.5\% \\
F1        & 62.9\% & 85.1\% & \textbf{87.3\%} \\
\hline
    \rowcolor[HTML]{DCDCDC}
\#Candidates & 4265 & 1519 & 821 \\
\hline\hline
\end{tabular}
}
\end{table}

\subsection{Results of RQ3: Ablation Study}
\label{sec:eval:rq3}

\begin{table}[!t]
    \centering
    \caption{Ablation Study of Major Components.}
    \label{tab:ablation_study}
    \renewcommand{\arraystretch}{1.4}
    \resizebox{\columnwidth}{!}{
    \begin{tabular}{c|ccc|c|c|c}
    \hline
    \hline
    \textbf{Configuration}
    & \multicolumn{3}{c|}{\textbf{Time-Shift}}
    & \textbf{Diverse-Family}
    & \textbf{Privileged-Benign}
    & \textbf{Overall} \\
    \cline{2-7}
    & \textbf{FNR} & \textbf{FPR} & \textbf{F1}
    & \textbf{FNR}
    & \textbf{FPR}
    & \textbf{F1} \\
    \hline

    \rowcolor[HTML]{CFE2F3}
    {w/o Multi-Signals}
    & 20.0\% & 1.7\% & 88.1\%
    & 33.7\% 
    & 7.4\%
    & 83.3\% \\

    {w/o Verification}
    & 9.8\% & 21.7\% & 85.1\%
    & 24.5\%
    & 47.4\%
    & 74.9\% \\

    \rowcolor[HTML]{CFE2F3}
    {w/o Context} 
    & 95.3\% & 4.4\% & 8.7\%
    & 67.5\%
    & 1.5\%
    & 24.7\% \\

    \hline
    \rowcolor[HTML]{DCDCDC}
    \textbf{Full Setting}
    & \textbf{11.7\%} & \textbf{3.3\%} & \textbf{92.2\%}
    & \textbf{20.9\%}
    & \textbf{13.0\%}
    & \textbf{87.4\%} \\
    
    \hline
    \hline
    \end{tabular}
    }
\end{table}

Table~\ref{tab:ablation_study} shows that every component contributes to detection performance. \emph{First}, without multiple signals to hypothesize from (\emph{w/o Multi-Signals}), recall drops: the Time-Shift FNR rises from 11.7\% to 20.0\% and the Diverse-Family FNR from 20.9\% to 33.7\%, because sensitive APIs alone miss behaviors that leave no sensitive-API trace, such as app uninstallation driven through the Android \texttt{Intent} mechanism with action \texttt{ACTION\_DELETE}. \emph{Second}, without verification (\emph{w/o Verification}), precision drops: the Time-Shift FPR rises from 3.3\% to 21.7\% and the Privileged-Benign FPR from 13.0\% to 47.4\%, since nothing forces the LLM's hypotheses to be backed by executable evidence and plausible but unsupported behaviors are flagged as malicious. \emph{Third}, without context-aware judgment (\emph{w/o Context}), recall collapses: the Time-Shift FNR rises from 11.7\% to 95.3\%, as the judge, though still given behaviors and their evidence, cannot separate a malicious behavior from legitimate app functionality. Each stage thus guards against a distinct failure, and recovered behavior alone is not enough to decide maliciousness.

\subsection{Results of RQ4: Effectiveness and Cost Across LLMs}
\label{sec:eval:rq4}


\begin{table}[!t]
    \centering
    \caption{Detection Performance under Different LLMs.}
    \label{tab:cross_models}
    \renewcommand{\arraystretch}{1.4}
    \resizebox{\columnwidth}{!}{
    \begin{tabular}{c|ccc|c|c|c}
    \hline
    \hline
    \textbf{Model}
    & \multicolumn{3}{c|}{\textbf{Time-Shift}}
    & \textbf{Diverse-Family}
    & \textbf{Privileged-Benign}
    & \textbf{Overall} \\
    \cline{2-7}
    & \textbf{FNR} & \textbf{FPR} & \textbf{F1}
    & \textbf{FNR}
    & \textbf{FPR}
    & \textbf{F1} \\
    \hline

    GPT-5.4-mini
    & 8.1\% & 9.0\% & 91.0\%
    & 19.0\%
    & 22.2\%
    & 84.4\% \\

    GPT-5.2
    & 20.0\% & 1.0\% & 88.3\%
    & 23.9\%
    & 7.4\%
    & 85.4\% \\

    \hline
    \rowcolor[HTML]{DCDCDC}
    \textbf{DeepSeek-V4-Pro}
    & \textbf{11.7\%} & \textbf{3.3\%} & \textbf{92.2\%}
    & \textbf{20.9\%}
    & \textbf{13.0\%}
    & \textbf{87.4\%} \\
    \hline
    \hline
    \end{tabular}
   
    }
\end{table}


Table~\ref{tab:cross_models} shows that all three models achieve overall F1 above 84\%, indicating that \tool maintains strong effectiveness across different LLMs.
The models exhibit distinct effectiveness trade-offs.
GPT-5.2 achieves the lowest FPR (1.0\% on time-shift and 7.4\% on privileged-benign) but also the highest FNR (20.0\%), indicating that it adopts a more conservative decision strategy.
GPT-5.4-mini shows the opposite pattern, achieving the highest recall but the highest FPR (22.2\% on privileged-benign), suggesting that it is more likely to classify ambiguous behaviors as malicious.
DeepSeek-V4-Pro balances these two extremes and achieves the best overall F1 (87.4\%).


Cost also varies substantially across models.
Function summarization always runs on the fixed lightweight model (DeepSeek-V4-Flash), so its cost stays at approximately \$0.56 per sample regardless of the detection model.
The remaining cost is \$0.04 with DeepSeek-V4-Pro, \$0.29 with GPT-5.4-mini, and \$0.67 with GPT-5.2.
Despite costing far more per judgment, GPT-5.2 does not exceed DeepSeek-V4-Pro's F1, making DeepSeek-V4-Pro the most cost-effective of the three.

\section{Threats to Validity}
\label{sec:discussion}
Our study has three threats to validity.
\emph{First}, our dataset may not represent all malware in the wild. To mitigate this, we collect 1,033 APKs spanning 2015 to 2026 and split them into three settings, each targeting a distinct dimension of diversity: temporal drift, behavioral coverage across 163 malware families, and functional coverage across 49 Google Play categories.
\emph{Second}, judging manually whether each recovered behavior is malicious is inherently subjective. To mitigate this, we ground RQ2 in public threat reports from recognized security organizations such as Kaspersky, ESET, and Cisco Talos, plus independent researchers; because such reports may omit implementation details or undocumented behaviors, our annotators also inspect the source code to confirm each behavior. We follow the dual-annotator protocol of \S\ref{sec:eval:rq2}, reaching Cohen's $\kappa$ of $0.88$ and $0.83$ for behavior identification and per-stage analysis.
\emph{Third}, \tool relies on an LLM to hypothesize, confirm, and judge behaviors, so its verdicts could vary with the choice of model and with nondeterministic decoding. To mitigate this, we decode at temperature 0 and take the majority verdict over three runs, and our cross-model evaluation (\S\ref{sec:eval:rq4}) shows that \tool stays effective across LLMs of different capability and cost.

\section{Related Work}
\label{sec:related}
\noindent
\textbf{Android Malware Detection.}
Android malware detection has evolved from hand-crafted static features~\cite{arp2014drebin,garcia2018lightweight, wu2021android,li2021robust} to structural program representations~\cite{onwuzurike2019mamadroid, wu2019malscan, zhang2020enhancing, he2022msdroid, zheng2024maskdroid} and LLM-based code analysis~\cite{zhao2025apppoet, li2025foredroid, qian2025lamd}.
Early feature-based detectors~\cite{arp2014drebin,wu2021android,li2021robust} extract permissions, API calls, and other static features, and train ML classifiers to identify malicious patterns.
Drebin~\cite{arp2014drebin} uses sparse static features with a linear SVM, while later methods such as XMal~\cite{wu2021android} and RAMDA~\cite{li2021robust} learn more discriminative or robust feature representations.
However, feature-based detectors capture individual features but not the structural relations among program elements.
Later detectors model API-call sequences or call-graph structure.
MaMaDroid~\cite{onwuzurike2019mamadroid} abstracts API calls to package-level states and models their transitions as Markov chains.
Graph-based detectors such as MalScan~\cite{wu2019malscan}, HomDroid~\cite{wu2021homdroid}, MsDroid~\cite{he2022msdroid}, and MaskDroid~\cite{zheng2024maskdroid} further use function call graphs or sensitive-API-centered subgraphs to improve detection and robustness.
Recent LLM-based detectors shift from learned representations to behavior-level reasoning.
ForeDroid~\cite{li2025foredroid} summarizes entry-to-sink API-call chains with an LLM, trains an OCSVM on the summaries, and uses the LLM for explanation.
LAMD~\cite{qian2025lamd} progressively summarizes multi-level backward slices around sensitive APIs to infer malicious behaviors in a training-free pipeline.
Despite these advances, existing detectors provide behavioral explanations without grounding them in connected code evidence. In contrast, \tool produces evidence-grounded behavior explanations together with the detection result.

\noindent
\textbf{Malicious Behavior Understanding.}
Prior work has also studied how security-relevant behaviors are implemented and interpreted in APKs.
Manual efforts such as Cao et al.~\cite{cao2022rotten} and MalRadar~\cite{wang2022malradar} provide high-quality behavior annotations for Android malware families, but require extensive expert reverse engineering.
ProMal~\cite{wu2025beyond} reduces manual effort by using an expert-built knowledge graph to interpret malware behaviors from API-level evidence.
However, it assumes a known malware as input, and its knowledge graph is constructed at a coarse granularity around APIs and parameters rather than complete behaviors.
Moreover, the knowledge graph is not automatically updated from newly analyzed malware.
InconPreter~\cite{yue2025s} analyzes UI-triggered API chains to explain risky behaviors and expose inconsistencies between claimed and implemented functionality. However, it remains API-chain-centric rather than recovering complete malicious behaviors.

\section{Conclusion}
\label{sec:conclusion}
We present \tool, an Android malware detection framework that recovers evidence-grounded malicious behaviors beyond binary detection. Across three challenging evaluation settings, \tool achieves 87.4\% overall F1, outperforming the strongest baseline by 18.6 percentage points. Its recovered behavior records match ground-truth malicious behaviors at 87.3\% F1, a substantial gain over the state-of-the-art baseline at 30.8\%. \tool remains effective across three LLMs, demonstrating that its behavior-oriented reasoning generalizes across different foundation models.





\bibliographystyle{IEEEtran}
\bibliography{main}

@String{Computing = "Computing" }

@String{Computer = "{IEEE} Computer" }

@inproceedings{pendlebury2019tesseract,
  title={$\{$TESSERACT$\}$: Eliminating experimental bias in malware classification across space and time},
  author={Pendlebury, Feargus and Pierazzi, Fabio and Jordaney, Roberto and Kinder, Johannes and Cavallaro, Lorenzo},
  booktitle={28th USENIX security symposium (USENIX Security 19)},
  pages={729--746},
  year={2019}
}

@ArtifactSoftware{R,
    title = {R: A Language and Environment for Statistical Computing},
    author = {{R Core Team}},
    organization = {R Foundation for Statistical Computing},
    address = {Vienna, Austria},
    year = {2019},
    url = {https://www.R-project.org/},
}

@inproceedings{wu2021homdroid,
  title={Homdroid: detecting android covert malware by social-network homophily analysis},
  author={Wu, Yueming and Zou, Deqing and Yang, Wei and Li, Xiang and Jin, Hai},
  booktitle={Proceedings of the 30th acm sigsoft international symposium on software testing and analysis},
  pages={216--229},
  year={2021}
}

@inproceedings{wu2019malscan,
  title={Malscan: Fast market-wide mobile malware scanning by social-network centrality analysis},
  author={Wu, Yueming and Li, Xiaodi and Zou, Deqing and Yang, Wei and Zhang, Xin and Jin, Hai},
  booktitle={2019 34th IEEE/ACM International Conference on Automated Software Engineering (ASE)},
  pages={139--150},
  year={2019},
  organization={IEEE}
}

@article{onwuzurike2019mamadroid,
  title={Mamadroid: Detecting android malware by building markov chains of behavioral models (extended version)},
  author={Onwuzurike, Lucky and Mariconti, Enrico and Andriotis, Panagiotis and Cristofaro, Emiliano De and Ross, Gordon and Stringhini, Gianluca},
  journal={ACM Transactions on Privacy and Security (TOPS)},
  volume={22},
  number={2},
  pages={1--34},
  year={2019},
  publisher={ACM New York, NY, USA}
}

@inproceedings{zhang2020enhancing,
  title={Enhancing state-of-the-art classifiers with api semantics to detect evolved android malware},
  author={Zhang, Xiaohan and Zhang, Yuan and Zhong, Ming and Ding, Daizong and Cao, Yinzhi and Zhang, Yukun and Zhang, Mi and Yang, Min},
  booktitle={Proceedings of the 2020 ACM SIGSAC Conference on Computer and Communications Security (CCS)},
  pages={757--770},
  year={2020}
}

@article{he2022msdroid,
  title={Msdroid: Identifying malicious snippets for android malware detection},
  author={He, Yiling and Liu, Yiping and Wu, Lei and Yang, Ziqi and Ren, Kui and Qin, Zhan},
  journal={IEEE Transactions on Dependable and Secure Computing},
  volume={20},
  number={3},
  pages={2025--2039},
  year={2022},
  publisher={IEEE}
}

@inproceedings{zheng2024maskdroid,
  title={MaskDroid: Robust Android malware detection with masked graph representations},
  author={Zheng, Jingnan and Liu, Jiahao and Zhang, An and Zeng, Jun and Yang, Ziqi and Liang, Zhenkai and Chua, Tat-Seng},
  booktitle={Proceedings of the 39th IEEE/ACM International Conference on Automated Software Engineering},
  pages={331--343},
  year={2024}
}

@inproceedings{qian2025lamd,
  title={Lamd: Context-driven android malware detection and classification with llms},
  author={Qian, Xingzhi and Zheng, Xinran and He, Yiling and Yang, Shuo and Cavallaro, Lorenzo},
  booktitle={2025 IEEE Security and Privacy Workshops (SPW)},
  pages={126--136},
  year={2025},
  organization={IEEE}
}

@inproceedings{li2025foredroid,
  title={ForeDroid: Scenario-Aware Analysis for Android Malware Detection and Explanation},
  author={Li, Jiaming and Chen, Sen and Wu, Chunlian and Zhang, Yuxin and Fan, Lingling},
  booktitle={Proceedings of the 2025 ACM SIGSAC Conference on Computer and Communications Security},
  pages={1379--1393},
  year={2025}
}

@inproceedings{gao2024comprehensive,
  title={A Comprehensive Study of Learning-based Android Malware Detectors under Challenging Environments},
  author={Gao, Cuiying and Huang, Gaozhun and Li, Heng and Wu, Bang and Wu, Yueming and Yuan, Wei},
  booktitle={Proceedings of the 46th IEEE/ACM International Conference on Software Engineering (ICSE)},
  pages={1--13},
  year={2024}
}

@inproceedings{li2023black,
  title={Black-box Adversarial Example Attack towards FCG Based Android Malware Detection under Incomplete Feature Information},
  author={Li, Heng and Cheng, Zhang and Wu, Bang and Yuan, Liheng and Gao, Cuiying and Yuan, Wei and Luo, Xiapu},
  booktitle={32nd USENIX Security Symposium (USENIX)},
  pages={1181--1198},
  year={2023}
}

@inproceedings{he2023efficient,
  title={Efficient query-based attack against ML-based Android malware detection under zero knowledge setting},
  author={He, Ping and Xia, Yifan and Zhang, Xuhong and Ji, Shouling},
  booktitle={Proceedings of the 2023 ACM SIGSAC Conference on Computer and Communications Security (CCS)},
  pages={90--104},
  year={2023}
}

@inproceedings{reimers2019sentence,
  title={Sentence-bert: Sentence embeddings using siamese bert-networks},
  author={Reimers, Nils and Gurevych, Iryna},
  booktitle={Proceedings of the 2019 conference on empirical methods in natural language processing and the 9th international joint conference on natural language processing (EMNLP-IJCNLP)},
  pages={3982--3992},
  year={2019}
}

@book{robertson2009probabilistic,
  title={The probabilistic relevance framework: BM25 and beyond},
  author={Robertson, Stephen and Zaragoza, Hugo},
  volume={4},
  year={2009},
  publisher={Now Publishers Inc}
}

@inproceedings{arp2014drebin,
  title={Drebin: Effective and explainable detection of android malware in your pocket.},
  author={Arp, Daniel and Spreitzenbarth, Michael and Hubner, Malte and Gascon, Hugo and Rieck, Konrad and Siemens, CERT},
  booktitle={Network and Distributed System Security Symposium (NDSS)},
  volume={14},
  pages={23--26},
  year={2014}
}

@article{garcia2018lightweight,
  title={Lightweight, obfuscation-resilient detection and family identification of android malware},
  author={Garcia, Joshua and Hammad, Mahmoud and Malek, Sam},
  journal={ACM Transactions on Software Engineering and Methodology (TOSEM)},
  volume={26},
  number={3},
  pages={1--29},
  year={2018},
  publisher={ACM New York, NY, USA}
}

@misc{virustotal,
  title        = {VirusTotal},
  author       = {VirusTotal},
  howpublished = {\url{https://www.virustotal.com}},
  note         = {Accessed: 2026-05-01}
}

@inproceedings{allix2016androzoo,
  title={Androzoo: Collecting millions of android apps for the research community},
  author={Allix, Kevin and Bissyand{\'e}, Tegawend{\'e} F and Klein, Jacques and Le Traon, Yves},
  booktitle={Proceedings of the 13th international conference on mining software repositories},
  pages={468--471},
  year={2016}
}

@inproceedings{li2021robust,
  title={Robust android malware detection against adversarial example attacks},
  author={Li, Heng and Zhou, Shiyao and Yuan, Wei and Luo, Xiapu and Gao, Cuiying and Chen, Shuiyan},
  booktitle={Proceedings of the Web Conference 2021},
  pages={3603--3612},
  year={2021}
}

@article{arzt2014flowdroid,
  title={Flowdroid: Precise context, flow, field, object-sensitive and lifecycle-aware taint analysis for android apps},
  author={Arzt, Steven and Rasthofer, Siegfried and Fritz, Christian and Bodden, Eric and Bartel, Alexandre and Klein, Jacques and Le Traon, Yves and Octeau, Damien and McDaniel, Patrick},
  journal={ACM sigplan notices},
  volume={49},
  number={6},
  pages={259--269},
  year={2014},
  publisher={ACM New York, NY, USA}
}

@inproceedings{yang2015appcontext,
  title={Appcontext: Differentiating malicious and benign mobile app behaviors using context},
  author={Yang, Wei and Xiao, Xusheng and Andow, Benjamin and Li, Sihan and Xie, Tao and Enck, William},
  booktitle={2015 IEEE/ACM 37th IEEE international conference on software engineering},
  volume={1},
  pages={303--313},
  year={2015},
  organization={IEEE}
}

@inproceedings{liu2022promal,
  title={ProMal: precise window transition graphs for android via synergy of program analysis and machine learning},
  author={Liu, Changlin and Wang, Hanlin and Liu, Tianming and Gu, Diandian and Ma, Yun and Wang, Haoyu and Xiao, Xusheng},
  booktitle={Proceedings of the 44th International Conference on Software Engineering},
  pages={1755--1767},
  year={2022}
}

@inproceedings{cao2022rotten,
  title={Rotten apples spoil the bunch: an anatomy of Google Play malware},
  author={Cao, Michael and Ahmed, Khaled and Rubin, Julia},
  booktitle={Proceedings of the 44th International Conference on Software Engineering},
  pages={1919--1931},
  year={2022}
}

@article{wang2022malradar,
  title={MalRadar: Demystifying android malware in the new era},
  author={Wang, Liu and Wang, Haoyu and He, Ren and Tao, Ran and Meng, Guozhu and Luo, Xiapu and Liu, Xuanzhe},
  journal={Proceedings of the ACM on Measurement and Analysis of Computing Systems},
  volume={6},
  number={2},
  pages={1--27},
  year={2022},
  publisher={ACM New York, NY, USA}
}

@inproceedings{ma2016libradar,
  title={Libradar: Fast and accurate detection of third-party libraries in android apps},
  author={Ma, Ziang and Wang, Haoyu and Guo, Yao and Chen, Xiangqun},
  booktitle={Proceedings of the 38th international conference on software engineering companion},
  pages={653--656},
  year={2016}
}

@inproceedings{yan2022iccbot,
  title={Iccbot: fragment-aware and context-sensitive icc resolution for android applications},
  author={Yan, Jiwei and Zhang, Shixin and Liu, Yepang and Yan, Jun and Zhang, Jian},
  booktitle={Proceedings of the ACM/IEEE 44th international conference on software engineering: companion proceedings},
  pages={105--109},
  year={2022}
}

@article{ruggia2025dark,
  title={The dark side of native code on android},
  author={Ruggia, Antonio and Possemato, Andrea and Dambra, Savino and Merlo, Alessio and Aonzo, Simone and Balzarotti, Davide},
  journal={ACM Transactions on Privacy and Security},
  volume={28},
  number={2},
  pages={1--33},
  year={2025},
  publisher={ACM New York, NY}
}

@inproceedings{samhi2022difuzer,
  title={Difuzer: Uncovering suspicious hidden sensitive operations in android apps},
  author={Samhi, Jordan and Li, Li and Bissyand{\'e}, Tegawend{\'e} F and Klein, Jacques},
  booktitle={Proceedings of the 44th International Conference on Software Engineering},
  pages={723--735},
  year={2022}
}

@inproceedings{li2015iccta,
  title={Iccta: Detecting inter-component privacy leaks in android apps},
  author={Li, Li and Bartel, Alexandre and Bissyand{\'e}, Tegawend{\'e} F and Klein, Jacques and Le Traon, Yves and Arzt, Steven and Rasthofer, Siegfried and Bodden, Eric and Octeau, Damien and McDaniel, Patrick},
  booktitle={2015 IEEE/ACM 37th IEEE International Conference on Software Engineering},
  volume={1},
  pages={280--291},
  year={2015},
  organization={IEEE}
}

@inproceedings{samhi2022jucify,
  title={Jucify: A step towards android code unification for enhanced static analysis},
  author={Samhi, Jordan and Gao, Jun and Daoudi, Nadia and Graux, Pierre and Hoyez, Henri and Sun, Xiaoyu and Allix, Kevin and Bissyand{\'e}, Tegawend{\'e} F and Klein, Jacques},
  booktitle={Proceedings of the 44th International Conference on Software Engineering},
  pages={1232--1244},
  year={2022}
}

@article{xi2024gnndroid,
  title={GNNDroid: Graph-Learning Based Malware Detection for Android Apps With Native Code},
  author={Xi, Ning and Zhang, Yuchen and Feng, Pengbin and Ma, Siqi and Ma, Jianfeng and Shen, Yulong and Yang, Yale},
  journal={IEEE Transactions on Dependable and Secure Computing},
  volume={22},
  number={2},
  pages={1460--1476},
  year={2024},
  publisher={IEEE}
}

@inproceedings{li2025ui,
  title={UI-CTX: Understanding UI Behaviors with Code Contexts for Mobile Applications.},
  author={Li, Jiawei and Liu, Jiahao and Mao, Jian and Zeng, Jun and Liang, Zhenkai},
  booktitle={NDSS},
  year={2025}
}

@inproceedings{yue2025s,
  title={What's Done Is Not What's Claimed: Detecting and Interpreting Inconsistencies in App Behaviors.},
  author={Yue, Chang and Chen, Kai and Guo, Zhixiu and Dai, Jun and Sun, Xiaoyan and Yang, Yi},
  booktitle={NDSS},
  year={2025}
}

@article{liu2026unraveling,
  title={Unraveling the Key of Machine Learning-based Android Malware Detection},
  author={Liu, Jiahao and Zeng, Jun and Pierazzi, Fabio and Yang, Ziqi and Cavallaro, Lorenzo and Liang, Zhenkai},
  journal={ACM Transactions on Software Engineering and Methodology},
  year={2026},
  publisher={ACM New York, NY}
}

@incollection{vallee2010soot,
  title={Soot: A Java bytecode optimization framework},
  author={Vall{\'e}e-Rai, Raja and Co, Phong and Gagnon, Etienne and Hendren, Laurie and Lam, Patrick and Sundaresan, Vijay},
  booktitle={CASCON first decade high impact papers},
  pages={214--224},
  year={2010}
}

@misc{irata_muha2xmad,
  title        = {Technical analysis of IRATA android malware},
  author       = {Muhammad Hasan Ali},
  year         = {2022},
  howpublished = {\url{https://muha2xmad.github.io/malware-analysis/irata/}},
  note         = {Accessed: 2026-05-01}
}

@misc{malwarevolumn,
  title        = {IT threat evolution in Q1 2026. Mobile statistics},
  author       = {{WeLiveSecurity, ESET}},
  year         = {2026},
  howpublished = {\url{https://securelist.com/malware-report-q1-2026-mobile-statistics/119819/}},
  note         = {Accessed: 2026-05-01}
}

@misc{malwarebazaar,
  author       = {{abuse.ch}},
  title        = {MalwareBazaar},
  year         = {2026},
  howpublished = {\url{https://bazaar.abuse.ch/}},
  note         = {Accessed: 2026-05-01}
}

@misc{ghidra,
  author       = {{National Security Agency}},
  title        = {Ghidra Software Reverse Engineering Framework},
  year         = {2026},
  howpublished = {\url{https://github.com/nationalsecurityagency/ghidra}},
  note         = {Accessed: 2026-05-01}
}

@inproceedings{xi2019deepintent,
  title={Deepintent: Deep icon-behavior learning for detecting intention-behavior discrepancy in mobile apps},
  author={Xi, Shengqu and Yang, Shao and Xiao, Xusheng and Yao, Yuan and Xiong, Yayuan and Xu, Fengyuan and Wang, Haoyu and Gao, Peng and Liu, Zhuotao and Xu, Feng and others},
  booktitle={Proceedings of the 2019 ACM SIGSAC Conference on Computer and Communications Security},
  pages={2421--2436},
  year={2019}
}

@article{song2025fcghunter,
  title={FCGHUNTER: Towards Evaluating Robustness of Graph-Based Android Malware Detection},
  author={Song, Shiwen and Xie, Xiaofei and Feng, Ruitao and Guo, Qi and Chen, Sen},
  journal={IEEE Transactions on Software Engineering},
  year={2025},
  publisher={IEEE}
}

@article{wu2021android,
  title={Why an android app is classified as malware: Toward malware classification interpretation},
  author={Wu, Bozhi and Chen, Sen and Gao, Cuiyun and Fan, Lingling and Liu, Yang and Wen, Weiping and Lyu, Michael R},
  journal={ACM Transactions on Software Engineering and Methodology (TOSEM)},
  volume={30},
  number={2},
  pages={1--29},
  year={2021},
  publisher={ACM New York, NY, USA}
}

@article{wu2025beyond,
  title={Beyond decision: Android malware description generation through profiling malicious behavior trajectory},
  author={Wu, Chunlian and Chen, Sen and Li, Jiaming and Chai, Renchao and Fan, Lingling and Xie, Xiaofei and Feng, Ruitao},
  journal={ACM transactions on software engineering and methodology},
  volume={34},
  number={7},
  pages={1--39},
  year={2025},
  publisher={ACM New York, NY}
}

@inproceedings{christodorescu2007mining,
  title={Mining specifications of malicious behavior},
  author={Christodorescu, Mihai and Jha, Somesh and Kruegel, Christopher},
  booktitle={Proceedings of the the 6th joint meeting of the European software engineering conference and the ACM SIGSOFT symposium on The foundations of software engineering},
  pages={5--14},
  year={2007}
}

@inproceedings{fredrikson2010synthesizing,
  title={Synthesizing near-optimal malware specifications from suspicious behaviors},
  author={Fredrikson, Matt and Jha, Somesh and Christodorescu, Mihai and Sailer, Reiner and Yan, Xifeng},
  booktitle={2010 IEEE Symposium on Security and Privacy},
  pages={45--60},
  year={2010},
  organization={IEEE}
}

@article{zhao2025apppoet,
  title={Apppoet: Large language model based android malware detection via multi-view prompt engineering},
  author={Zhao, Wenxiang and Wu, Juntao and Meng, Zhaoyi},
  journal={Expert Systems with Applications},
  volume={262},
  pages={125546},
  year={2025},
  publisher={Elsevier}
}
\end{document}